\documentclass[twocolumn]{revtex4-2}
\usepackage[dvipsnames]{xcolor}
\usepackage{amsmath}
\usepackage{amssymb}
\usepackage{amsthm}
\usepackage{mathtools}
\usepackage{graphicx}
\usepackage{enumitem}

\definecolor{lgray}{gray}{0.35}
\usepackage[colorlinks=true,urlcolor=lgray,linkcolor=lgray,
citecolor=lgray,pdfpagelabels=true,hypertexnames=true,
plainpages=false,naturalnames=false]{hyperref}
\usepackage{graphicx}

\newcommand{\Vcat}{V_{\text{cat}}^{(\delta)}}

\newcommand{\beq}{\begin{equation}}
	\newcommand{\eeq}{\end{equation}}
\newcommand{\beqnn}{\begin{equation*}}
	\newcommand{\eeqnn}{\end{equation*}}
\newcommand{\ber}{\begin{array}}
	\newcommand{\eer}{\end{array}}
\newcommand{\ena}{\end{eqnarray}}
\newcommand{\beqa}{\begin{eqnarray}}
\newcommand{\eeqa}{\end{eqnarray}}
\newcommand{\bea}{\begin{eqnarray}}
\newcommand{\eea}{\end{eqnarray}}

\begin{document}

\title{On cat-human interaction from the viewpoint of physics: an equation of motion}

\author{Anxo Biasi}
\email{anxo.biasi@gmail.com}
\affiliation{Laboratoire de Physique de l'Ecole Normale Sup\'erieure ENS Universit\'e PSL,
	CNRS, Sorbonne Universit\'e, Universit\'e de Paris, F-75005 Paris, France.
}

\begin{abstract}	
	\noindent This paper provides an enjoyable example through which several concepts of classical mechanics can be understood. We introduce an equation that models the motion of a cat in the presence of a person. The cat is considered as a point particle moving in a potential induced by the person. We demonstrate that this approach to the problem reproduces characteristic behaviors of these curious animals. For instance, the fact that cats do not typically come when they are called, or that they remain longer on the lap of their favorite person; even ``zoomies" are reproduced (cats randomly run back and forth across the house). We use this model problem to explore topics of current research such as stochastic equations and periodically driven systems. The pedagogical value of this work and its potential use in teaching are discussed.
\end{abstract}

\maketitle


\section{Introduction}

Cats are such charming animals that people frequently immortalize them in pictures and videos that are shared with friends, relatives, or even the world. The popularity of cats has grown to the point that they gained their own spot on the Internet, occupying a significant portion of the content on social media. A quick search reveals numerous cat accounts that have amassed millions of views and followers. This astonishing popularity is partly due to their unique behavior. But what causes these animals to behave in their characteristic way? This paper offers a physical perspective on the subject. In the low-energy limit, when cats are calmer, we have identified seven common behaviors of cats and built an equation that qualitatively reproduces them. To our knowledge, this work is the first to model features of cat behavior with an equation, although other cat features have already been explored by physicists, such as their incredible ability to land on their feet \cite{FallingCats1,FallingCats2,FallingCats3}.

We will model the cat as a point particle and see what energy potentials can explain the collection of behaviors cats display in the presence of a person, which we list in section~\ref{sec:Cats_phenomenology}.
In section~\ref{sec:Cats_equation}, the model equation of cat motion is introduced, and in section~\ref{sec:P1-P5}, we explore how potential minima and friction combine to reproduce some of the characteristic behaviors cats display. In section~\ref{sec:Zoomies}, the stochastic version of the equation of motion is introduced to show that a random forcing term can be added to reproduce random behaviors of cats. In section~\ref{sec:Purring}, purring is explained in our model by the presence of an external periodic forcing term, that improves the stability of equilibrium points. 

This work combines classical mechanics with an intriguing story that may be used to familiarize students with concepts like equilibrium points, potential barriers, friction, or external forcing. These concepts may be, of course, introduced just by using traditional point particle setups. However, it is common that students
in initial courses find difficulties in the abstract
nature of physics \cite{DifficultiesPhysics}. In particular, unfamiliar setups (e.g. a system of particles) hinder the visualization of key ideas. This paper's approach may help students better understand mechanics concepts thanks to three elements: 1) the low level of abstraction demanded by our setup (a cat and a person), 2) the ease of visualization of this real world scenario, 3) the curiosity aroused in both expert and general audiences. For these reasons, we believe this work may turn into valuable material in introductory classical mechanics courses.


\section{Cat behavior}
\label{sec:Cats_phenomenology}

The inspiration to understand cat behavior from the viewpoint of physics came from our cat. Since she arrived, we started to observe peculiar behaviors in her interaction with us. After some time, we noticed repetitive patterns, making us view her motions as those of a point particle (a physicist never rests!). We realized that her dynamics are rather simple, with well-established equilibrium points that are stable or unstable under external stimulus. Then, we started thinking about whether some of her characteristic behaviors could be modeled by an equation. We here focus on a simple scenario of the cat-human interaction: {\em a single cat in the presence of a single person at rest}. The dynamics we observed are the following:

\begin{itemize}
	\item {\bf P1:} Cats commonly rest keeping some distance from people.
	
	\item {\bf P2:} When cats rest on a person (on the lap, belly, back, etc), minimal stimuli may provoke them to leave that position (a fly, an imperceptible sound, a $\beta$-decay of an atom in a neighboring galaxy, etc). The intensity of perturbations needed to produce the departure depends on the attachment cats have to the person they are resting on.
	 
	\item {\bf P3:}  When cats are petted by people, they move back and forth in an oscillatory motion.

	\item {\bf P4:} When cats are called by people, they seldom respond.
	
	\item {\bf P5:} When cats decide to approach the person who calls them, they often get distracted on the way and do not reach the person.
	
	\item {\bf P6:} At night, cats randomly run back and forth across the house. These episodes are called ``zoomies" (and we invite the reader to search for some videos on the Internet!).
	
	\item {\bf P7:} Cats purr (emit a soft and vibrating sound) when they like being petted by a person.
	
\end{itemize}

Points {\bf P1-P7} have been extracted from the observation of the quotidian life of our cat, discussions with friends about their own cats, the common lore, the lore on the Internet, and specialized material on cat behavior \cite{ReviewCatHumanInteraction,ReviewCats}. Therefore, they are not universal and some cats may display a weaker version of some of them. Note that this is a theoretical work intended for physics education and that no experiments on animals have been conducted.


\section{The cat's equation}
\label{sec:Cats_equation}

\begin{figure}[t]
	\includegraphics[width = 0.97 \columnwidth]{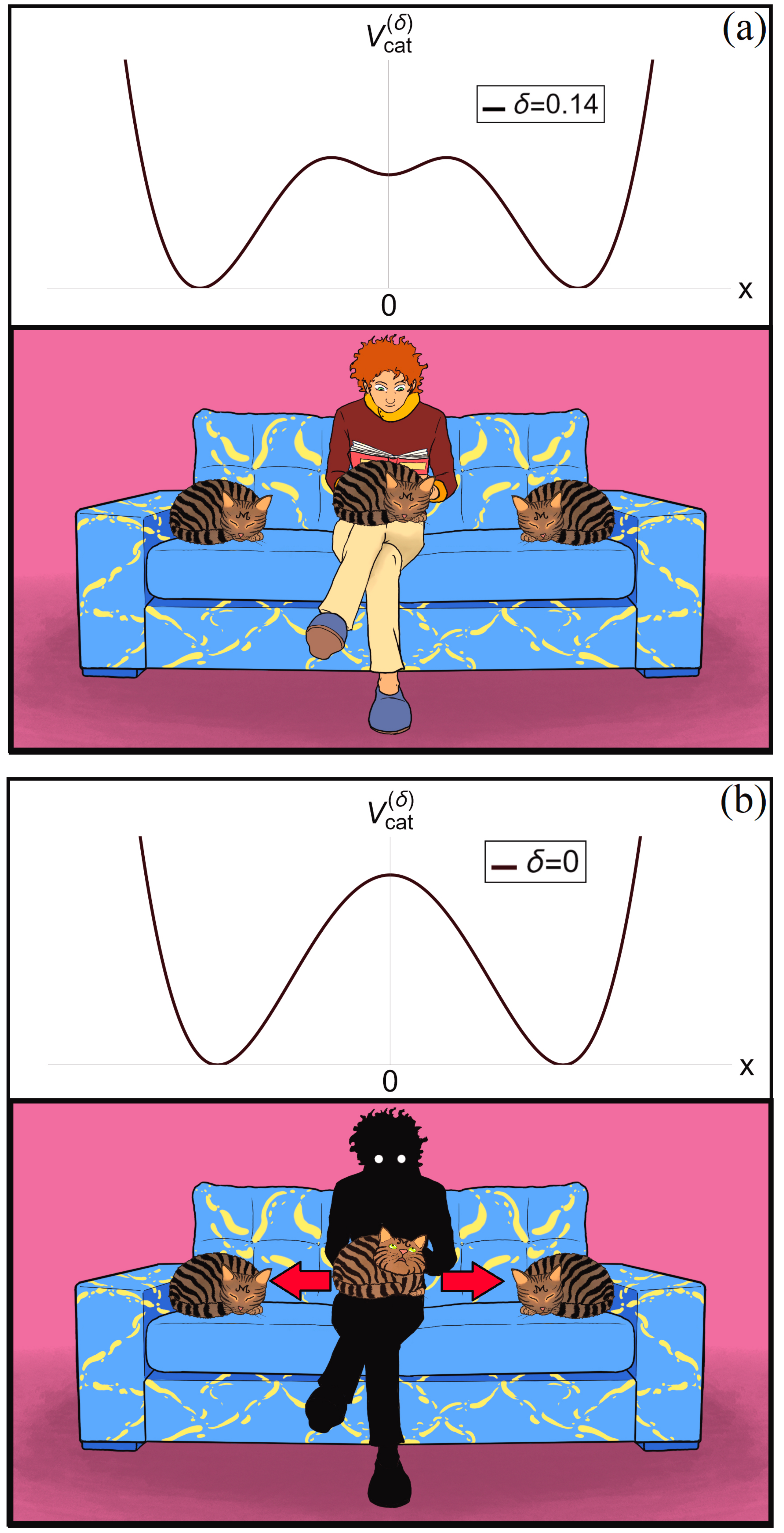}
	\caption{Interpretation of the cat-human interaction potential $\Vcat$. In (a) the cat is attached to the person ($\delta = 0.14$), while in (b) the person is a complete stranger to the cat ($\delta=0$). In both cases the cat is illustrated at rest at three equilibrium points of $\Vcat$.}
	\label{fig:Sofa_cat-human}
\end{figure} 

Our working hypothesis is that cats behave as if they perceive a force around a person. As a first approximation, we consider that their dynamics are modeled as a point particle obeying Newton's mechanics in the presence of an external potential $\Vcat$ (induced by a person at rest) and a friction term
\beq
m\frac{d^2x}{dt^2} = -\frac{d \Vcat(x)}{dx} - \epsilon \frac{dx}{dt},
\label{eq:EOM}
\eeq
where $x(t)\in\mathbb{R}$ represents the cat's position at time $t$ with respect to the person placed at $x=0$, $m>0$ is the cat's mass, and $\epsilon>0$ is the friction coefficient with values that depend on each cat. In the next section, we will see that the following potential reproduces behaviors {\bf P1}-{\bf P5}, giving a good idea of how cats perceive the presence of people (see Fig.~\ref{fig:Sofa_cat-human}):
\beq
\Vcat(x) = g (x^2-1)x^2\frac{x^2-\delta}{x^2+\delta}.
\label{eq:Vcat}
\eeq
This form is based on a rational function for the control it provides over the equilibrium points, enabling us to manage their number, relative position, and stability. As we shall see, these are the key ingredients, together with friction, to reproduce behaviors {\bf P1}-{\bf P5}.
 $g>0$ is the coupling constant (set to $g=1$ from now on to simplify the notation, but this does not alter the qualitative picture of the model), and $\delta\in[0,1]$ reflects the attachment the cat has to the person. For $\delta = 0$ the point $x=0$ (the person's position) is unstable: the cat has no attachment to the person. On the other hand, for $\delta>0$ the point $x=0$ is stable, as Fig.~\ref{fig:Sofa_cat-human} shows. Larger values of $\delta$ are associated with stronger bonds. For $\delta = 1$ the minimum at the origin is as deep as the off-centered ones, indicating that the cat has a very strong attachment to the (lucky!) person.

The asymptotic growth of the potential, $\Vcat(x\to \pm\infty)\to\infty$, is  used to keep the cat close to the person, modeling that cat and person are confined in the same room. Potentials with this kind of asymptotic growth are often encountered in research, as experiments in a laboratory are usually localized in space. For instance, they are used in cold atom experiments \cite{ColdAtoms}, or for modeling light propagation in some optical fibers \cite{OpticalFibers}.

The friction term in Eq.~(\ref{eq:EOM}) is needed to reduce the energy,
\beq
E = \frac{m}{2}\left(\frac{dx}{dt}\right)^2 + \Vcat(x), \qquad \frac{dE}{dt} = -\epsilon \left(\frac{dx}{dt}\right)^2,
\label{eq:Energy_eq}
\eeq
otherwise, the cat does not converge to resting positions after periods of activity. Note that, in order to oppose the cat's motion, we need $\epsilon\geq0$ and friction must be proportional to an odd power of the velocity. Finally, let us note that the cat moves in a 3-dimensional space. However, since the most important parameter is the distance between the cat and the person, we assume the cat moves along a line, which further simplifies the analysis.

\section{Reproducing cat behaviors {\bf P1}-{\bf P5}}
\label{sec:P1-P5}

In this section, we show that the cat's equation (Eq.~(\ref{eq:EOM})) captures qualitatively the phenomenological observations {\bf P1}-{\bf P5}.

	{\bf P1}: ``{\em Cats commonly rest keeping some distance from people}". This behavior is captured by the off-centered global minima present in $\Vcat$, as illustrated in Fig.~\ref{fig:Sofa_cat-human}. The cat may start at many positions with different velocities but will end up at the minima due to the friction term. On most occasions, the final position will be a global minimum, specially for a weak bond between the cat and the person (small $\delta$). However, if the bond is strong enough, the person's position competes with the off-centered minima for the cat's preference to rest. This scenario is captured by our model when $\delta\to 1$, as $x=0$ becomes an additional global minimum.

	{\bf P2}: ``{\em  When cats rest on a person (...), minimal stimuli may provoke them to leave that position (...). The intensity of perturbations needed to produce the departure depends on the attachment cats have to the person they are resting on.}" The first part of this statement is reproduced by the equilibrium point of the potential at $x=0$ (the person's position) because the cat may rest on the person $x(t)=\dot{x}(t)=0$. The second part of the statement is captured by the dependence of the potential on $\delta$. From the second derivative at the origin
	\beq
		\frac{d^2\Vcat}{dx^2}\bigg{|}_{x=0} = \begin{cases}
			-2 & \text{if } \delta = 0,\\
				2 & \text{if } \delta > 0,
		\end{cases}
	\eeq
	 we see that $x=0$ is unstable for $\delta = 0$, indicating a departure of the cat from the person under arbitrarily small perturbations. On the other hand, for $\delta>0$, this position is stable and stronger stimulus is needed to detach the cat from the person as $\delta$ grows. In section~\ref{sec:Purring}, we will see that other phenomena have stability regions that depend on a continuous parameter such as $\delta$, for instance, in systems subject to a time-periodic forcing, e.g. Kapitza's pendulum \cite{Kapitza,BookLandauMechanics}.

\begin{figure}[t!]
		\includegraphics[width = 0.85 \columnwidth]{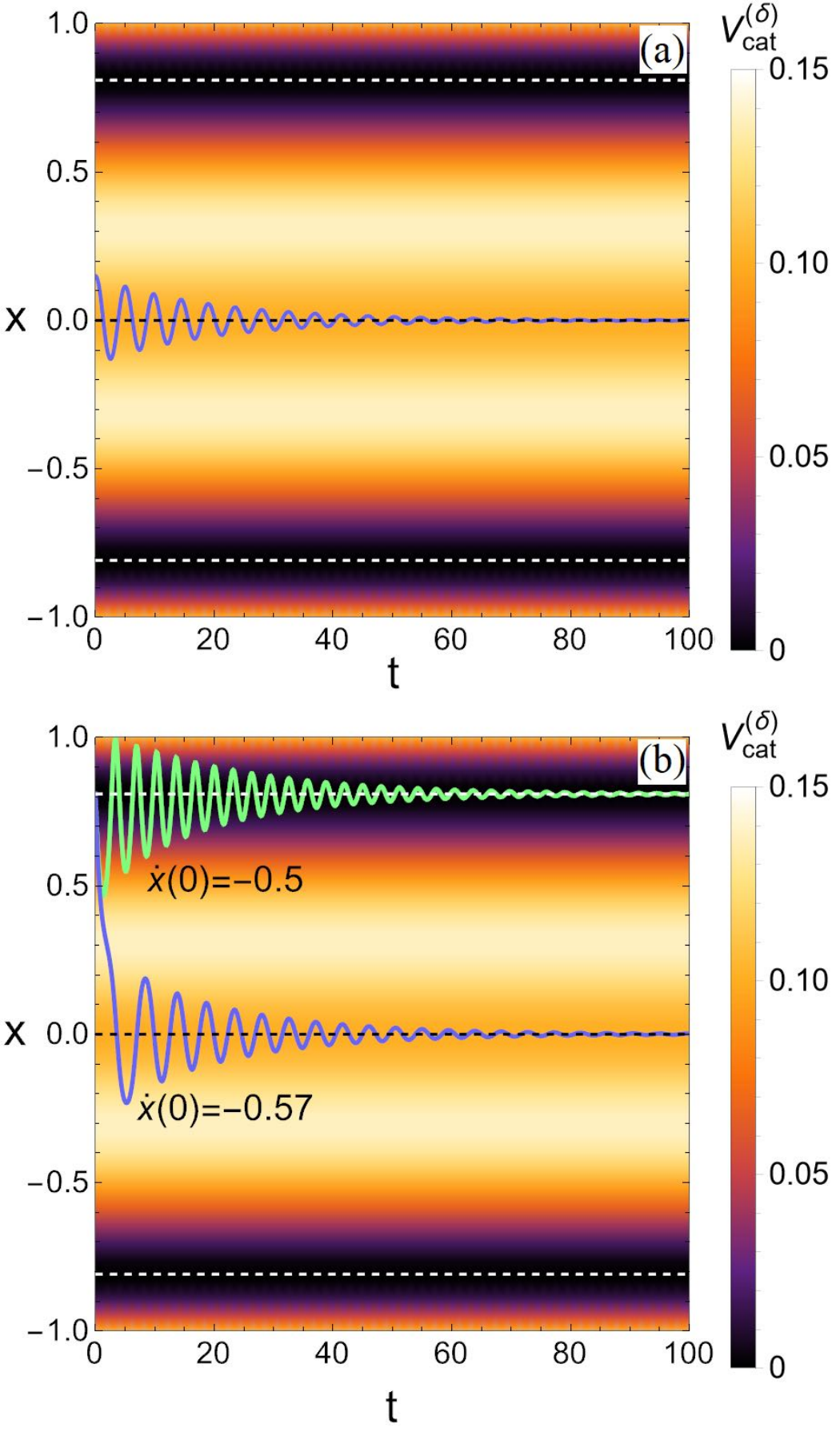}
		
		\caption{Space-time representation of cat's trajectories with the values of the potential shown in color, white dashed lines mark the off-centered global minima and the black one the local minimum at $x=0$. (a) Trajectory initially close to the person ($x(t=0)=0.15$, $\dot{x}(t=0)=0$). (b) Trajectories from the minimum $(x(t=0)=x_{\min}\approx 0.81)$ for two different initial velocities (see the label next to each curve). The cat-human bond has been set to $\delta = 0.25$, the friction to $\epsilon = 0.1$, and the mass to $m=1$.}
		\label{fig:P3_P4_P5}
	\end{figure}

	{\bf P3:} ``{\em  When cats are petted by people, they move back and forth in an oscillatory motion.}" This effect is also reproduced by the region of stability around $x=0$ (for $\delta>0$). As Fig.~\ref{fig:P3_P4_P5}(a) illustrates, when cats are calm and close to a person (low kinetic energy), they perform small amplitude oscillations around the person, converging to the static state $x=\dot{x}=0$ (resting), thanks to the friction term in Eq.~(\ref{eq:EOM}). As we explain in section~\ref{sec:Purring}, the cat may display these stable oscillations even when petted by a stranger ($\delta=0$), but for this, a new ingredient must be added (purring).  Note that the motion observed in Fig.~\ref{fig:P3_P4_P5}(a) is the one described by a damped oscillator, as can for instance be observed in the realization of the Cavendish experiment \cite{Cavendish}.

	Observations {\bf P4} and  {\bf P5} are explained in terms of the potential barrier between an off-centered minimum and the person ($x=0$), see Fig.~\ref{fig:Sofa_cat-human}. {\bf P4:} ``{\em When cats are called by people, they seldom respond.}" The act of being called is modeled as an impulse of the cat toward the person, which results in an increment of the kinetic energy. This energy injection may be sufficient or insufficient to overcome the potential barrier, as illustrated by the blue (darker) and green (lighter) trajectories in Fig.~\ref{fig:P3_P4_P5}(b), respectively. For an  insufficient energy, the cat converges again to the resting position after some time, $x=x_{\min}$. {\bf P4} refers to the fact that cats often find the call to be so unmotivated that their interest is rapidly dissipated. In our model, this represents a very small increase of energy that is consumed by friction before the cat moves far. We see that cats need to experience a strong impulse to come to a person. This fact is often overlooked, provoking a feeling of frustration when the cat does not react to our call. This has generated the  wrong idea that cats are ``selfish" animals as Ref.~\cite{ReviewCats} discusses, whereas cats just happen to have a stronger inner damping mechanism, compared to, say, dogs.
	
	When the cat is given a sufficiently strong stimulus (impulse) to approach the person, we observe {\bf P5:} ``{\em  When cats decide to approach the person who calls them, they often get distracted on the way and do not reach the person.}" This observation is also explained in terms of the damping mechanism described above, as illustrated by  the green (lighter) trajectory in Fig.~\ref{fig:P3_P4_P5}(b). It is interesting to discuss the role played by the cat's mass  in this phenomenon. Since an external stimulus results in an injection of kinetic energy ($E_K$), the velocity the cats acquire decreases with increasing the mass according to $\dot{x} = \sqrt{2E_K/m}$. It clearly aligns with the observation that light cats, such as kittens, exhibit energetic movements and react to any stimulus, while heavier cats, such as elderly or overfed cats, do not show the same enthusiasm. However, one must consider carefully the relation between the cat's mass ($m$) and friction ($\epsilon$) to avoid unrealistic scenarios. To investigate this question, we write the cat's equation (Eq.~(\ref{eq:EOM})) in its dimensionless form 
	\beq
		\frac{d^2\tilde{x}}{d\tau^2} = -\frac{d \Vcat(\tilde{x})}{d\tilde{x}} - \tilde{\epsilon} \frac{d\tilde{x}}{d\tau},
	\label{eq:EOM_dimensionless}
	\eeq
	where $\tau = t /\sqrt{m}$, $\tilde{x}(\tau) = x(t/\sqrt{m})$, and $\tilde{\epsilon} = \epsilon /\sqrt{m}$. The energy variation takes the form:
	\beq
		\frac{dE}{d\tau} = - \tilde{\epsilon} \left(\frac{d\tilde{x}}{d\tau}\right)^2,
	\eeq
	indicating that $\epsilon$ must depend on $m$ to model cats of different masses. To illustrate this fact let us assume that $\epsilon$ is independent of $m$. In this situation, heavy cats would consume less energy than lighter ones because $\tilde{\epsilon}$ decreases with $m$, i.e., elderly and overfed cats would not get tired, while kittens would be quickly exhausted (a world upside down!). This unrealistic situation is avoided by considering that the friction coefficient changes with the cat's mass. If $\epsilon$ is proportional to $\sqrt{m}$, then cats would describe the same trajectory $\tilde{x}(\tau)$ independently of their mass. When the trajectory is expressed back in time $t = \tau  \sqrt{m}$, cats perform the same activity (they visit the same positions $x$) but require a shorter or longer time depending on their mass. We still believe this is not fully realistic, as kittens usually perform activities that  are not displayed by elderly cats. We then conclude that the friction coefficient must grow faster than  $\sqrt{m}$, e.g., $\epsilon \propto m$, restoring the natural order of life where elderly cats (larger $\tilde{\epsilon}$) get exhausted more quickly than kittens (smaller $\tilde{\epsilon}$).


\section{Zoomies as a stochastic process}
\label{sec:Zoomies}

Zoomies ({\bf P6}), also known as frenetic random activity periods (FRAPs), are phases during which cats have an excess of energy that makes them suddenly run from place to place, usually at night. In our model, we assume that zoomies happen when the cat randomly moves between minima of $\Vcat$, representing long runs from one side of a room to the other. However, these random processes cannot be reproduced by our deterministic equation (Eq.~(\ref{eq:EOM})). If the cat has enough energy to move between minima of the potential, Eq.~(\ref{eq:EOM}) predicts that this will not be a random process but will translate into repeated damped oscillations. At some point, the decreasing energy will not be sufficient to overcome the potential barrier for the cat to move to other minima. The best one can expect from deterministic systems is great sensitivity to initial conditions, in which case small variations in the conditions may lead to completely different trajectories (the so-called butterfly effect \cite{Butterfly}). Even so, the energy would be consumed by friction, eventually impeding the displacement between minima. For these reasons, our deterministic setup requires an external forcing that randomly injects and extracts energy, producing a different trajectory each time the equation is solved for the same initial conditions. Namely, we turn the cat's equation into a stochastic equation \cite{BookStochasticEuations}. This approach is useful in systems that are subject to high uncertainty, such as the movement of suspended particles in a liquid (Brownian motion), which are subject to unpredictable collisions with the liquid particles \cite{Feynman}. 

\begin{figure}[t]
	\includegraphics[width = 0.9\columnwidth]{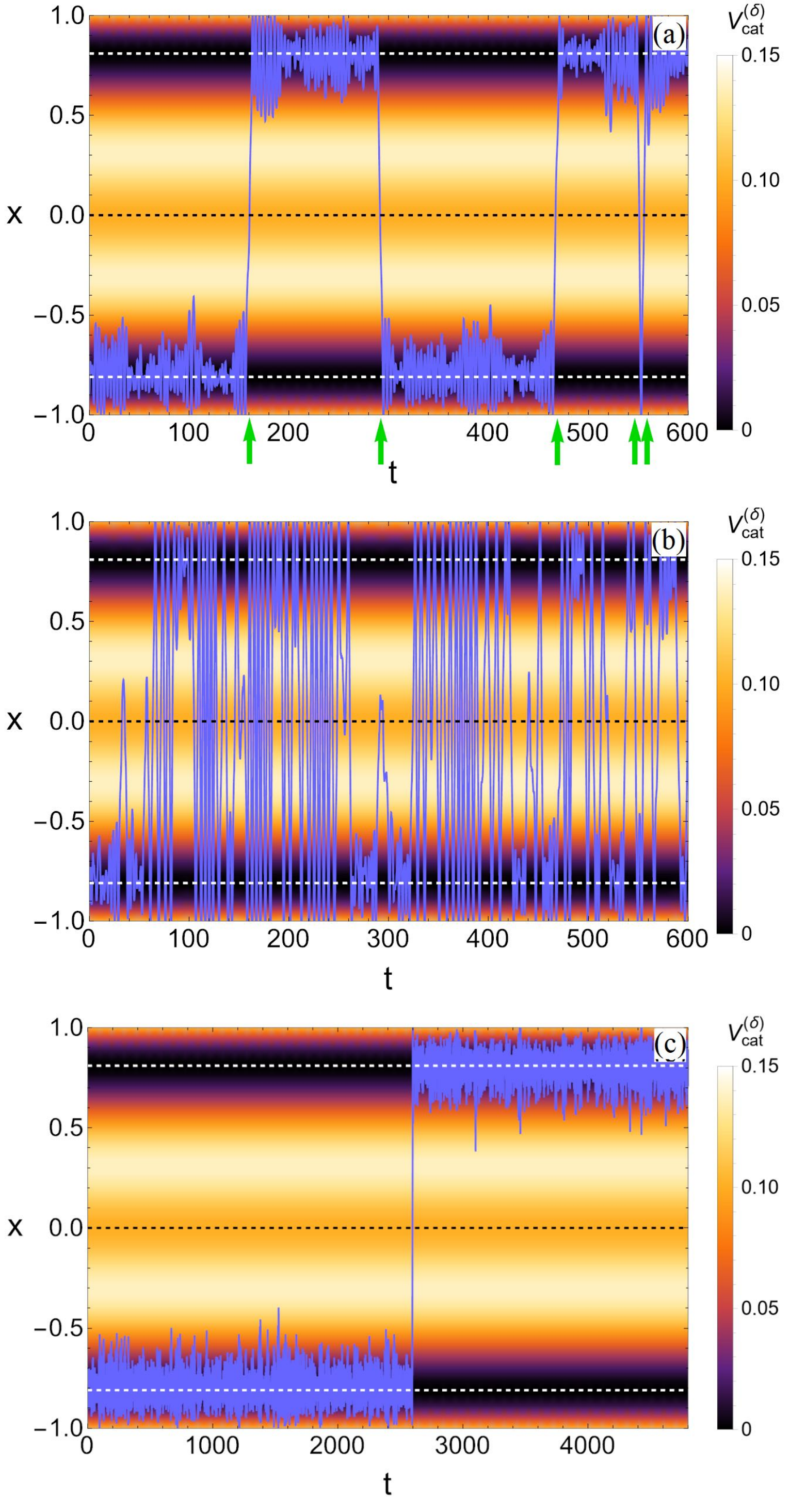}
	
	\caption{Space-time evolution of the stochastic cat's equation (\ref{eq:Stochastic_EOM}). The colored background represents the potential $\Vcat$, the solid line the position of the cat, the white dashed lines the off-centered potential minima and the black dashed line the origin. Green arrows in plot (a) mark zoomies, when the cat randomly runs from one minimum to another. In this plot, we used the following parameters $(x(t=0),\dot{x}(t=0),\delta,\epsilon,m,\sigma) = (-0.9,0,0.25,0.1,1,0.1)$. Plots (b) and (c) illustrate how friction and forcing have a great impact on the production of zoomies by using $(\epsilon,\sigma)=(0.05,0.2)$ and $(\epsilon,\sigma)=(0.2,0.095)$, respectively. Note that plot (c) presents an evolution eight times longer than (a) and (b).}
	\label{fig:Zoomies}
\end{figure}

The stochastic cat's equation has the form
\beq
	m\frac{d^2x}{dt^2} = -\frac{d \Vcat(x)}{dx} - \epsilon \frac{dx}{dt} + \sigma f(t)
\label{eq:Stochastic_EOM}
\eeq
where $\sigma$ is a constant and $f(t)$ the external random forcing. This equation can be solved by numerical integration, using the Euler-Maruyama method \cite{StochasticNumerics}, which is the Euler method adapted to stochastic equations \cite{BookStochasticEuations}. As $f(t)$, we use white noise \cite{BookStochasticEuations}, a common phenomenon in physics, which is found in the aforementioned Brownian motion, or in more applied scenarios as in the dynamics of financial markets (e.g. Black–Scholes model \cite{Review_Black-Scholes,BookStochasticEuations}). As Fig.~\ref{fig:Zoomies}(a) illustrates, Eq.~(\ref{eq:Stochastic_EOM}) produces zoomies: the cat may suddenly run from one equilibrium point to another, remain there for some time, and again randomly go back to the previous equilibrium point. The probability of producing zoomies in a given period of time depends on the values of the friction $\epsilon$ and the forcing $\sigma$. This allows us to adjust the model to the particularities of each cat. For instance, kittens constantly exhibit these periods, corresponding to lower friction and higher forcing (Fig.~\ref{fig:Zoomies}(b)) than elderly cats, which rarely display such activity (Fig.~\ref{fig:Zoomies}(c)).

\section{Purring: a stabilization mechanism}
 \label{sec:Purring}
 
 {\bf P7:} ``{\em Cats purr (emit a soft and vibrating sound) when they like being petted by a person.}" In this section, we propose that purring is a stabilization mechanism. First, we point out that when a cat is being petted and begins purring, people generally get the urge to keep petting the cat, strengthening in this way the stability of the process. The second reason is the analogy with Kapitza's pendulum \cite{Kapitza,Kapitza2}. While the inverted pendulum is normally unstable, it may become stable by introducing vibrations (small amplitude high-frequency oscillations) in the vertical direction. Vibrations then work as a stabilization mechanism for unstable equilibrium points. This is exactly the model we propose for purring: that vibrations strengthen the effective bond between cat and human.

 \begin{figure}[t!]
 	\includegraphics[width = 0.95 \columnwidth]{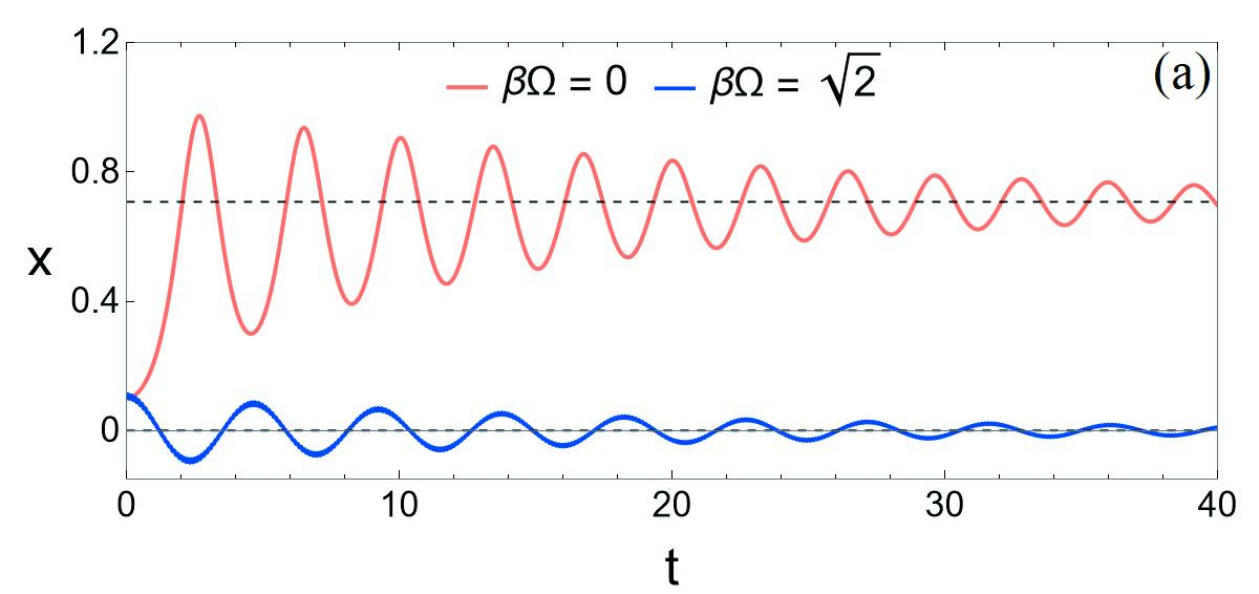}
 	
 	\includegraphics[width = 0.95 \columnwidth]{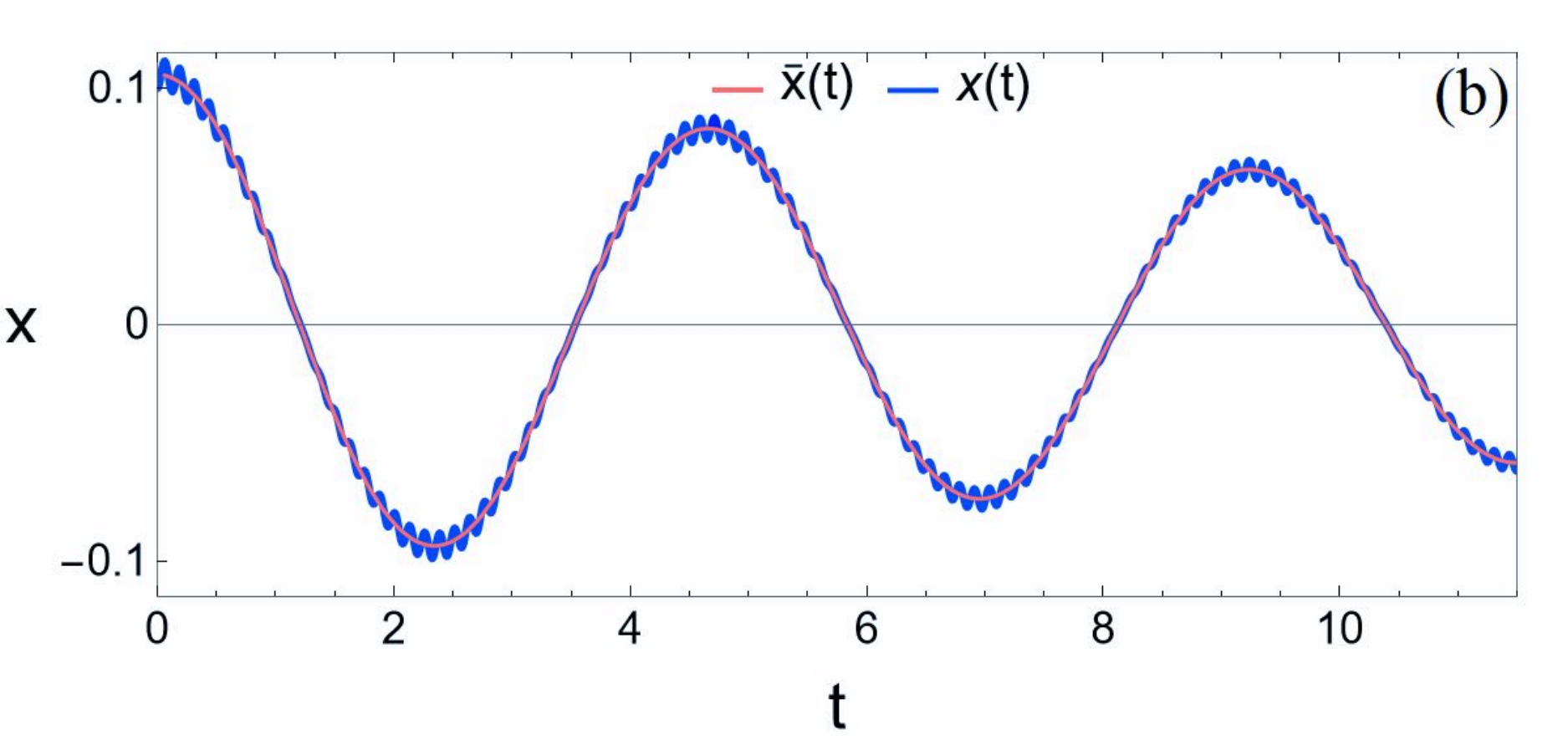}
 	
 	\includegraphics[width = 0.95 \columnwidth]{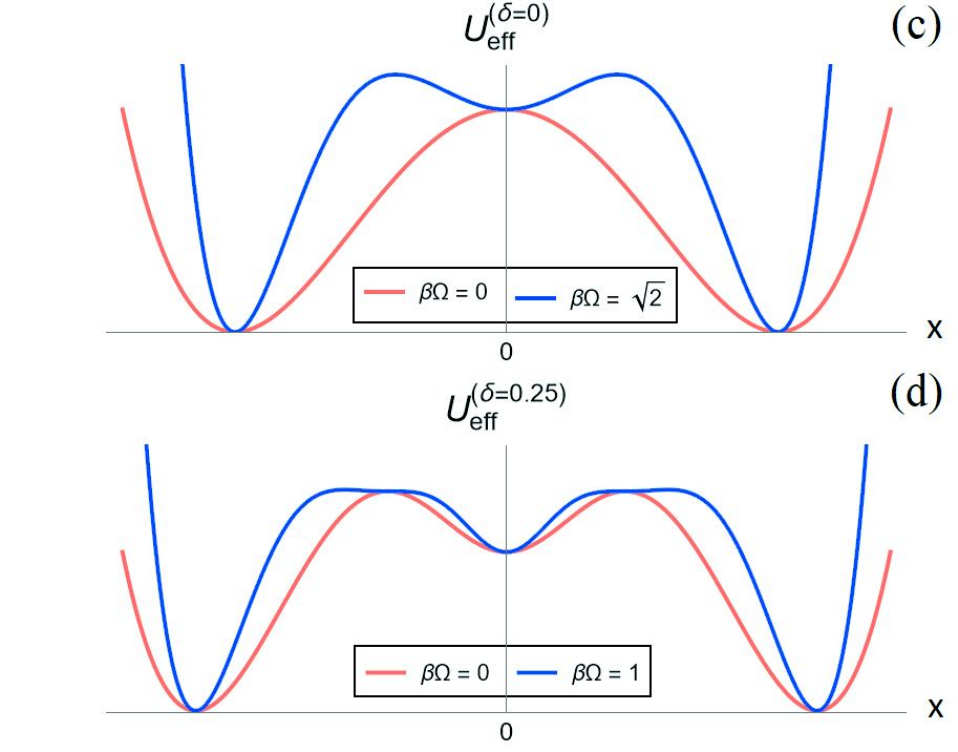}
 	
 	\caption{(a) Comparison between trajectories as given by  the cat's equation with and without external vibration, both initiated with the same initial conditions ($x(t=0)=0.1$, $\dot{x}(t=0)=0$), $\delta=0$, and $\epsilon=0.1$. Equilibrium points are marked by black dashed lines. (b) A closer view of the trajectory subject to external vibrations shown in plot (a) and its average motion. (c) Potential $\Vcat$ vs effective potential $	U_{\text{eff}}$ for $\delta=0$. (d) Potential $\Vcat$ vs effective potential $	U_{\text{eff}}$ for $\delta=0.25$. We fixed $m=1$ in these plots.}
 	\label{fig:Ueff}
 \end{figure}

 Noting that cats vibrate when they purr (as can be checked by placing a hand on their back), we introduce an external vibrating forcing in the cat's equation (\ref{eq:EOM}) to mimic the effect
 \beq
 	m\frac{d^2x}{dt^2} = -\frac{d \Vcat(x)}{dx} - \epsilon \frac{dx}{dt} + \beta \Omega^2 G(x) \cos(\Omega t),
 	\label{eq:EOM_purring}
 \eeq
 where $G(x)$ is an unconstrained function for now, and $\beta$ and $\Omega$ are the amplitude and frequency of the vibrations induced in the cat, respectively. Driving terms of this kind model a particle moving in an oscillating field in time or subject to periodic forcing \cite{BookLandauMechanics}. The typical example is Kapitza's pendulum \cite{Kapitza,Kapitza2}, where $G(\theta)$ is proportional to $\frac{dV_g}{d\theta}$, with $V_g$ being the gravitational potential and replacing $x$ by the angular coordinate $\theta$ \cite{Kapitza2}. We choose $\beta \ll 1$, and $\Omega \gg 1$ with $\beta \Omega \sim \mathcal{O}(1)$ in  order to reproduce the small amplitude high-frequency vibrations that are observed when purring. The goal is to determine the conditions that $G(x)$ must satisfy in order to replicate the effect of purring on the cat-human bond.
 
  The equation contains high-frequency oscillatory terms superposed to the original potential $\Vcat$. This means that the solutions get contributions at two time scales, $t\sim\mathcal{O}(\Omega^{-1})$ associated with the fast vibrations, and $t\sim\mathcal{O}(1)$ associated with the average motion of the cat, as we show by numerically solving the equation (Fig.~\ref{fig:Ueff}(b)). The latter motion may be obtained by averaging the trajectory over a period of fast oscillations:
 \beq
 \bar{x}(t) \equiv \frac{\Omega}{2\pi} \int_{t-\frac{\pi}{\Omega}}^{t+\frac{\pi}{\Omega}} x(s)ds.
 \label{eq:averaging}
 \eeq
 Decomposing the dynamics in slow and fast motions and using an averaging method (see appendix~\ref{sec:Appendix_averanging} for details), the average (or slow) motion of $x(t)$, denoted by $\bar{x}(t)$, is approximated by the trajectory of a cat submitted to a time-independent effective potential of the form
\beq
	U_{\text{eff}}(\bar{x}) = \Vcat(\bar{x}) + \frac{(\beta \Omega)^2}{4m} G(\bar{x})^2.
	\label{eq:Ueff_purring_cat}
\eeq
The equilibrium points of this potential now depend on $\Vcat(\bar{x})$ and $G(\bar{x})$, indicating that the amplitude of the external forcing plays an important role. An analysis at the origin, the point relevant to us, reveals that $G(x=0)G'(x=0)=0$ and $G'(x=0)^2 + G(x=0) G''(x=0)>0$ are required to maintain the person's position as an equilibrium point and to enhance its stability during purring. As an example, $G = \frac{d\Vcat}{dx}$ satisfies these conditions, maintaining the original equilibrium points, but their stability/instability may change depending on the product $\beta \Omega$,
\beq
	\frac{d^2U_{\text{eff}}}{d\bar{x}^2}\bigg{|}_{\bar{x}=0} = \begin{cases}
		-2+2\frac{(\beta\Omega)^2}{m} & \text{if } \delta =0, \vspace{0.2cm}\\
		2+2\frac{(\beta\Omega)^2}{m} & \text{if } \delta >0.
		\end{cases}
\eeq
The unstable point $x=0$ for $\delta = 0$ turns into a stable one when $\beta\Omega>m^{1/2}$ (see Fig.~\ref{fig:Ueff}(c)),  indicating that introducing external vibrations in our model has similar effects on the stability of $x=0$ as increasing $\delta$. For $\delta>0$, $x=0$ is already stable, but the stability is strengthened with the growth of $\beta\Omega$, in the sense that $U_{\text{eff}}$ grows from the origin faster than $\Vcat$ (see Fig.~\ref{fig:Ueff}(d)). Thus, for the same energy, displacements from the origin have a smaller amplitude when forcing is introduced, indicating  that the cat has a stronger effective attachment to the person. These observations  suggest that cats purr to temporarily strengthen the bond with humans, using vibrations to evoke a feeling that makes people want to keep petting them.

Similar stability in periodically driven systems is observed in other contexts. In quantum many-body physics, the topic has been under intense research in the last decades \cite{FloqueReview1,FloqueReview2}, both theoretically and experimentally. It has been shown that periodic driving may be particularly useful to manipulate systems, for instance, it allows to reach new phases of matter such as topological insulators \cite{FloquetInsulator}.


\section{Conclusions}
\label{sec:Conclusions}

We have explored cat-human interaction from the viewpoint of physics. We first identified seven characteristic dynamics a cat displays in the presence of a person at rest ({\bf P1}-{\bf P7}). Considering the cat as a point particle moving in a potential induced by the person, we presented an equation of motion (Eq.~(\ref{eq:EOM})) that reproduces these behaviors. With this first approximation to the cat-human interaction, we have built a qualitative picture that reflects the way cats perceive the presence of people, something that can be visualized by simply plotting a potential. This work establishes an arena to explore other features of cat-human interactions from the viewpoint of physics, such as the presence of several people, or even addressing new scenarios such as cat-cat, dog-dog, or dog-human interactions.

The cat-human interaction model we have introduced is intended to be used in introductory courses in classical mechanics to familiarize students with notions such as equilibrium points, potential barriers, friction, or external forcing. The model has three aspects that would benefit the exposition of concepts: 1) a low level of abstraction, 2) a collection of real-world dynamics (cat behavior) that facilitates visualization of results, and 3) a curious story that calls the attention of students. For these reasons, we believe that the model would help students to assimilate concepts while they see, at the same time, that physics may be used in an enjoyable way.


\section*{Acknowledgments} I am grateful to my cat for being my source of inspiration. I also thank Paloma~Calder\'on~Bustillo, Brad~Cownden, Piotr~Bizo\'n, and \'Angel~Paredes for useful discussions. During the development of this work, I have been supported by the LabEx ENS-ICFP: ANR-10-LABX-0010/ANR-10-IDEX-0001-02 PSL*.


\section*{Disclaimer}

This work is for educational purposes in physics and must not be understood as research on cat or human behavior. Furthermore, no experiments on animal or human subjects have been conducted to develop this work.


\section*{Conflict of Interest}

The author has no conflicts to disclose.

\appendix


\section{Averaging approximation}
\label{sec:Appendix_averanging}

We describe the procedure \cite{BookLandauMechanics} followed in section~\ref{sec:Purring} to construct an approximation of the solution to the cat's equation under external vibrating forcing (\ref{eq:EOM_purring}):
\beq
m\frac{d^2x}{dt^2} = -\frac{d \Vcat(x)}{dx} - \epsilon \frac{dx}{dt} + \beta \Omega^2 G(x) \cos(\Omega t),
\label{eq:Appendix_EOM_purring}
\eeq
where $G(x)$ is an unconstrained function for now, $\beta\ll1$ and $\Omega\gg1$ with $\beta \Omega \sim \mathcal{O}(1)$. We proceed perturbatively by introducing the small parameter $\xi\ll 1$ such that
\beq
\beta = \xi \tilde{\beta}, \quad \text{and} \quad \Omega = \xi^{-1}\tilde{\Omega},
\eeq
with $\tilde{\beta},\ \tilde{\Omega} \sim \mathcal{O}(1)$. The solution is then decomposed in the form
\beq
x(t) =  \xi z(t/\xi) + y(t),
\label{eq:y_z_decomposition}
\eeq
with $y,\ z\sim \mathcal{O}(1)$. The first term represents the leading part of the fast evolution (i.e., small-amplitude high-frequency oscillations), while $y$ contains the slow evolution and subleading terms of the fast motion. This separation will allow us to first obtain an expression for $z$, followed by an effective equation for the average motion of $y$. Note that highly oscillatory terms in $y$ cannot be neglected at this point because they will appear in the equation for $y$; that is why we need an averaging. Note as well that differentiation of $\xi z(t/\xi)$ increases its order ($\tau := t/\xi$)
\beq
\xi \frac{dz(t/\xi)}{dt} = \frac{dz(\tau)}{d\tau} \sim \mathcal{O}(1), 
\eeq 
\beq
\xi \frac{d^2z(t/\xi)}{dt^2} = \xi^{-1}\frac{d^2z(\tau)}{d\tau^2} \sim \mathcal{O}\left(\xi^{-1}\right).
\eeq 
Plugging (\ref{eq:y_z_decomposition}) into (\ref{eq:Appendix_EOM_purring}) and gathering terms of the same order in $\xi$, we get that the leading order equation is
\beq
m\frac{d^2z}{d\tau^2} = \tilde{\beta} \tilde{\Omega}^2 G(y) \cos(\tilde\Omega \tau),
\label{eq:EOM_purring_z}
\eeq
while the next order equation is
\beq
m\frac{d^2y}{dt^2} = - \frac{d\Vcat(y)}{dy} - \epsilon\left(\frac{dy}{dt}+\frac{dz}{d\tau}\right) + \frac{dG(y)}{dy}z \tilde{\beta}\tilde{\Omega}^2 \cos(\tilde{\Omega} \tau).
\label{eq:EOM_purring_y}
\eeq
The solution to (\ref{eq:EOM_purring_z}) can be approximated at leading order by
\beq
z(\tau) \approx  - m^{-1}\tilde{\beta} G(y) \cos(\tilde\Omega \tau).
\label{eq:z_tau}
\eeq 
We used the fact that $y(t)$ has two motions: subleading fast oscillations, which we neglected in the expression, and the slow motion that evolves on the time scale $t\sim\mathcal{O}(1)$. The latter remains practically constant when the equation is integrated over $\tau$ ($t\sim\mathcal{O}(\xi)$), namely, in an interval of order $\xi$. Therefore, one may perform the integration of (\ref{eq:EOM_purring_z}) considering $G(y)$ as a constant. See Ref.~\cite{Murdock} for the details of this approximation. The intuitive idea comes from the fact that, to leading order,
\beq
	\int_{t_0}^{t_0 + \xi \tau} f(t) \sin\left(\tilde{\Omega} \frac{t}{\xi}\right) dt \approx f(t_0) \int_{t_0}^{t_0 + \xi \tau} \sin\left(\tilde{\Omega} \frac{t}{\xi}\right) dt
\eeq
when $f(t)$ is a function evolving over time scales $t\sim\mathcal{O}(1)$.

Plugging the expression for $z(\tau)$ in (\ref{eq:EOM_purring_y}) one gets a right-hand side with terms that evolve with a typical time scale $t$ but also highly oscillatory terms in $\tau$. To get the slow evolution of $y(t)$, we average the motion over periods of the fast oscillations similar to what we did in Eq.~(\ref{eq:averaging}), getting $\left<\sin(\tilde{\Omega}\tau)\right> = 0$ and $\left<\cos^2(\tilde{\Omega}\tau)\right> = 1/2$. With that, an equation for the average motion $\bar{y}(t)= \left<y(t)\right>$ is obtained:
\beq
m\frac{d^2\bar{y}}{dt^2} = -\frac{d\Vcat(\bar{y})}{dy}  - \frac{(\tilde{\beta}\tilde{\Omega})^2}{2m} G(\bar{y})\frac{dG(\bar{y})}{dy} -\epsilon\frac{d\bar{y}}{dt},
\eeq 
from which we infer that the effective potential has the form
\beq
U_{\text{eff}}(\bar{y}) = \Vcat(\bar{y}) + \frac{(\beta \Omega)^2}{4m} G(\bar{y})^2.
\eeq



\end{document}